\documentclass[10pt, conference]{IEEEtran}
\usepackage{amsfonts}
\usepackage{amsmath, amssymb}
\usepackage{graphicx}
\usepackage{epsfig}
\usepackage{color}
\newtheorem{thm}{Theorem}

\newtheorem{lem}{Lemma}

\IEEEoverridecommandlockouts

\begin{document}
\title{Capacity of Symmetric K-User Gaussian Very Strong Interference Channels}
\author{Sriram Sridharan, Amin Jafarian, Sriram Vishwanath and Syed. A. Jafar
\thanks{S. Sridharan, A. Jafarian and S. Vishwanath are with the Wireless Networking and Communications Group, Department of Electrical and Computer Engineering, University of Texas at Austin, Austin, TX - 78712 (email: sridhara@ece.utexas.edu; jafarian@ece.utexas.edu; sriram@ece.utexas.edu). S. Sridharan, A. Jafarian and S. Vishwanath are supported in part by National Science Foundation
grants NSF CCF-0448181, NSF CCF-0552741, NSF CNS-0615061, and NSF
CNS-0626903, THECB ARP and the Army Research Office YIP.}%
\thanks{S. A. Jafar is with the Department of Electrical and Computer Science, University of California at Irvine, Irvine, CA 96297 (email : syed@ece.uci.edu). S. A. Jafar is supported by ONR Young Investigator Award N00014-08-1-0872.}%
}
\maketitle
\begin{abstract}
This paper studies a symmetric $K$ user Gaussian interference channel with $K$ transmitters and $K$ receivers. A ``very strong" interference regime is derived for this channel setup. A ``very strong" interference regime is one where the capacity region of the interference channel is the same as the capacity region of the channel with no interference. In this regime, the interference can be perfectly canceled by all the receivers without incurring any rate penalties. A ``very strong" interference condition for an example symmetric $K$ user deterministic interference channel is also presented.
\end{abstract}
\section{Introduction}
Determining the capacity region of the Gaussian interference channel has been a long standing open problem. Exact capacity results are known only for some special classes of interference channels such as the two user ``very strong" interference channel in \cite{CarleialVeryStrongInterference} and the two user strong interference channel \cite{HanKobayashi1981, Sato1981}. Recent results on the capacity region of the two user Gaussian interference channel include - the characterization of the capacity region to within one bit per channel use in \cite{EtkinTseWang2007} and determining the sum capacity for the mixed interference regime and the very weak interference regime \cite{MotahariKhandani, ShangKramerChen2007, AnnapureddyVeeravalli}. For interference networks with more than $2$ users, capacity approximations within $o(\log(\mbox{SNR}))$ (also known as degrees of freedom) have been found for time-varying (or frequency-selective) channels with coefficients drawn from a continuous distribution \cite{CadambeJafarInt}. Capacity results are also available for multiuser extensions of the ``very weak interference'' scenario \cite{MotahariKhandani, ShangKramerChen2007, AnnapureddyVeeravalli} and for certain specific channel coefficient values, such as the toy examples in \cite{CadambeJafarInt}. In \cite{BreslerParekhTse}, the authors approximately characterize the capacity region of many-to-one and one-to-many Gaussian interference channels using abstractions to deterministic interference channels.

\begin{figure}
\centering\label{fig:lem1}
\includegraphics[width=1.8in]{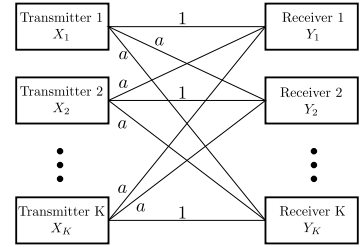}
\caption{Channel Model for the Symmetric $K$ user Interference Channel}
\label{fig:system model}
\end{figure}

In this paper, we study the $K$ user symmetric Gaussian interference channel (see Figure \ref{fig:system model}) and derive a ``very strong" interference regime for this channel. A ``very strong" interference regime is one where the capacity region of the interference channel is the same as the capacity region of the channel with no interference. In this regime, the interference can be perfectly canceled by all the receivers without incurring any associated rate penalties. We also present a ``very strong" interference condition for an example symmetric $K$ user deterministic interference channel (see Figure \ref{fig:detint}). The main tool used in this paper in deriving the ``very strong" interference condition is lattice coding, where the transmitted codewords are lattice points. Lattice and other structured coding techniques have been used recently to derive clever achievable schemes for several classes of networks. Some relevant results include \cite{BreslerParekhTse}--\nocite{ ErezZamir, ErezLitsynZamir, NamElGamal, KrithivasanPradhan, NazerGastparAllerton2007, NazerGastpar, Loeliger}\cite{UrbankeRimoldi}.

The rest of the paper is organized as follows: We describe the channel model in Section II. In Section III, we summarize the ``very strong" interference conditions for the two user Gaussian interference channel and present our main results for the $K$ user interference channel. We describe notations and preliminaries on lattices in Section IV. The proof of the main result for the three user interference channel is presented in Section V. We conclude in Section VI.

\section{Channel Model}
We consider the $K$ user interference channel, consisting of $K$ transmitters, $K$ receivers, and $K$ independent messages, where message $W_{k}$ originates at transmitter $k$ and is intended for receiver $k,\forall k\in \mathcal{K}\triangleq\{1,2,\cdots,K\}$. For simplicity of exposition, in this paper we consider only the symmetric (Gaussian) interference channel model. We employ lattice coding at the transmitters and choose lattices such that the interference lattices align themselves at each receiver. This idea can be generalized to a much wider class of non-symmetric channels. The symmetric channel model is described as follows.
\begin{eqnarray*}
Y_{j}(i) = X_{j}(i)+a\sum_{k=1,k\neq j}^K X_{k}(i)+Z_{j}(i), ~~\forall j\in\mathcal{K}
\end{eqnarray*}
where at the $i^{th}$ channel use, $Y_{j}(i)$ is the received signal at the $j^{th}$ receiver, $X_{k}(i)$ is the transmitted signal at the $k^{th}$ transmitter, $Z_{j}(i)$ is the zero mean unit variance additive white Gaussian noise (AWGN) at receiver $j$. All direct channels are normalized to unity, while all cross channels take the same value $a$, which is constant across channel uses. The channel inputs are subject to the transmit power constraint:
\begin{eqnarray}
\frac{1}{n}\sum_{i=1}^n\mbox{E}\left[{X_k(i)}^2\right]\leq P, ~~~\forall k\in\mathcal{K}.
\end{eqnarray}
Achievable rates, probability of error and capacity are defined in the Shannon sense.
\section{Very Strong Interference Condition}
\subsection{2 User IC}
Carleial \cite{CarleialVeryStrongInterference} showed that interference is not harmful when it is very strong, because the interfering signal can be decoded without any rate penalty for either the desired or the interfering user's message. For the symmetric channel described above, if $K=2$ and
\begin{eqnarray}
\frac{1}{2}\log\left(1+\frac{a^2P}{1+P}\right)\geq\frac{1}{2}\log\left(1+P\right),
\end{eqnarray}
then interference can be decoded first while treating the desired signal as noise and without limiting the rate of the interfering user's message. This gives us the very strong interference condition as:
\begin{eqnarray}
a^2\geq 1+P.
\end{eqnarray}
Each user achieves a rate $R=\frac{1}{2}\log\left(1+P\right)$ which is his individual capacity in the absence of interference. For our purpose, this is also the defining property of ``very strong interference'' - i.e.,  a $K$ user fully connected (all channel coefficients are non-zero) interference channel is called a ``very strong interference'' channel if every user achieves a rate equal to his individual capacity in the absence of all interference.

\subsection{$K$ User IC - Decoding Interference}
One simple extension of the very strong interference condition for the symmetric $K$ user interference channel is readily obtained as follows. If
\begin{eqnarray}
\frac{1}{K-1}\times\frac{1}{2}\log\left(1+\frac{(K-1)a^2P}{1+P}\right)\geq\frac{1}{2}\log\left(1+P\right),
\end{eqnarray}
then it is easily seen that each user can first jointly decode all interfering signals while treating his desired signal as noise and then subtract all interference from his received signal to achieve his interference-free capacity. This gives us the following ``very strong interference'' condition:
\begin{eqnarray}
a^2\geq \frac{((1+P)^{K-1}-1)(1+P)}{(K-1)P}.\label{eqn:strongdecode}
\end{eqnarray}
Condition (\ref{eqn:strongdecode}) shows that each user can achieve his individual interference-free capacity if the strength of the interference scales exponentially with the number of users. As we show in this paper, this condition can be tightened quite significantly. The following example formulated in terms of the deterministic channel illustrates the key insights.
\subsection{Very Strong Interference on the Deterministic Channel}
\begin{figure}
\centering
\includegraphics[width=2in]{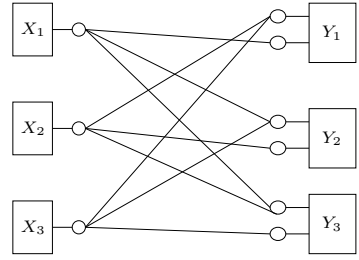}
\caption{Deterministic Channel Example}
\label{fig:detint}
\end{figure}
Fig. \ref{fig:detint} is the deterministic channel model (as proposed by \cite{BreslerParekhTse}) for a $3$ user fully connected interference channel. In this example, each user achieves a rate equal to the capacity that he would achieve in the absence of all interference.  Note that with all three users transmitting at capacity, a receiver is able to decode the desired message but cannot decode any of the two interferers. However, each receiver is able to decode the \emph{sum} of the codewords sent by the interfering users. For example, receiver $1$ cannot decode the messages $W_{2}, W_{3}$ but it can decode the sum of the interfering codewords $X_{2}+X_{3}$.

Note that the example illustrated in Fig. \ref{fig:detint} can be extended to any number of users. In the terminology of generalized degrees of freedom \cite{BreslerParekhTse} the ``very strong interference'' condition for this symmetric deterministic channel can be stated as:
\begin{eqnarray}
\frac{\log(\mbox{INR})}{\log(\mbox{SNR})}&\geq&2.
\end{eqnarray}
Since $\mbox{INR}=Pa^2$ and $\mbox{SNR}=P$, the example suggests a very strong interference condition of the form $a^2\geq P+o(P)$ for all $K$, instead of the exponential increase with $K$ evident in the RHS of (\ref{eqn:strongdecode}). These insights are most relevant for our main result - a very strong interference condition for the $K$ user Gaussian interference channel presented in the next section.
\subsection{$K$ User IC - Aligning Interference}
In this section we use lattice codes to align interference at each receiver in such a way that the sum of the interfering codewords can be decoded, without requiring the decodability of the messages carried by the interfering signals. Relaxing the message decodability constraint produces a much tighter ``very strong interference'' condition for the $K$ user symmetric interference channel. Lattice codes have previously been used in \cite{BreslerParekhTse} for interference alignment on the \emph{many-to-one} and \emph{one-to-many} interference channels, leading to capacity characterizations within a fixed number of bits per channel use for these channels. However, since we are interested in fully connected interference networks, several key aspects of the lattice code constructions in this section are unique to our setup. We present our main result below:
\begin{thm}\label{thm : main result}
For a $K$ user symmetric Gaussian interference channel, if the channel gain $a$ satisfies
\begin{equation}\label{eqn:tight constraint}
a^2 \geq \frac{(P+1)^2}{P},
\end{equation}
the capacity region of the channel, denoted by $\mathcal{C}_{ap}$ is given
\begin{eqnarray}\label{eqn:capacity}
\mathcal{C}_{ap} = \left\{\begin{array}{l}(R_1, \ldots, R_k) : \vspace{0.2cm}\\
\qquad R_k \leq \frac{1}{2}\log(1+P) \ \ \forall\ k \in \mathcal{K}\end{array}\right\}.
\end{eqnarray}
\end{thm}
In the rest of the paper, we prove this result for the three user interference channel $(K=3)$. The proof technique used here can readily be extended for any $K$. 

The region described by (\ref{eqn:capacity}) is an outer bound on the capacity region for a three user interference channel for any value of $a$. This is because $1/2 \log(1+P)$ is the maximum rate achieved by any user when there is no interference.
To show that the region described by (\ref{eqn:capacity}) for $K = 3$ is achievable under ``very strong interference" given by (\ref{eqn:tight constraint}), we show that the symmetric rate point $\left(\frac{1}{2}\log(1+P), \frac{1}{2}\log(1+P), \frac{1}{2}\log(1+P)\right)$ is achievable when (\ref{eqn:tight constraint}) is satisfied. The transmitters use lattice coding to encode their messages, while the receivers first decode the total interference and then decode their message after canceling all the interference. We describe some preliminaries on lattices in Section \ref{Sec:LatticeCoding} and also present some results on lattice codes derived in \cite{Loeliger} and \cite{UrbankeRimoldi} that will be used in the proof of Theorem \ref{thm : main result}. We then present the proof for achievability for the three user symmetric Gaussian interference channel under ``very strong interference" in Section \ref{Sec:Achievability}. 

Note that the ``very strong" interference condition for the $K$ user symmetric Gaussian interference channel is different from the condition for the two user case given by $a^2 \geq P + 1$. In fact, we have the following result for $a^2 \geq P + 1$ for the $K$ user symmetric Gaussian interference channel.

\begin{thm}\label{thm : side result}
For a $K$ user symmetric Gaussian interference channel, if the channel gain $a$ satisfies $a^2 \geq P+1$, then each user can achieve a rate of $\frac{1}{2}\log(P)$. Hence, for $a^2 \geq P+1$, each user achieves within half a bit per channel use of his maximum possible rate
\end{thm}
The idea behind the proof of Theorem 2 is described in Section \ref{Sec:Achievability}.

\section{Lattice Preliminaries and Notations}\label{Sec:LatticeCoding}
A lattice $\Lambda$ of dimension $n$ is a discrete subset of $\mathbb{R}^n$ described by
\begin{equation}
\Lambda = \{\lambda = G\mathbf{x} : \mathbf{x} \in \mathbb{Z}^n\},
\end{equation}
where $G$ is the generator matrix that defines the lattice $\Lambda$. Let $\Omega_{\Lambda}$ and $V_{\Lambda}$ denote the fundamental Voronoi region of lattice $\Lambda$ and the volume of $\Omega_{\Lambda}$ respectively.
We will drop the subscript in the Voronoi region and will refer to it as just $\Omega$.
In this paper, we consider lattices generated using construction A described below (as used in \cite{Loeliger}).

For any positive integer $p$, $\mathbb{Z}_p$ denotes the integers modulo $p$. Let $g : \mathbb{Z}^n \rightarrow \mathbb{Z}_p^n$ denote the componentwise modulo-$p$ operation. Let $\Lambda_C$ denote a lattice of the form
\begin{displaymath}
\Lambda_C = \left\{v \in \mathbb{Z}^n : g(v) \in C\right\}
\end{displaymath}
where $C$ is a linear code over $\mathbb{Z}_p$ (This is referred to as Construction A). In fact, we will actually consider scaled mod-$p$ lattices, i.e., lattices of the form $\gamma \Lambda_C = \left\{\gamma v : v \in \Lambda_C\right\}$ for some $\gamma \in \mathbb{R}$. The fundamental volume of such a lattice is
\begin{displaymath}
V_{\gamma \Lambda_C} = \gamma^n p^{n-k}.
\end{displaymath}

A set $\mathcal{B}$ of linear $(n,k)$ codes over $\mathbb{Z}_p$ is balanced if every nonzero element of $\mathbb{Z}_p^n$ is contained in the same number of codes from $\mathcal{B}$. Let $\mathcal{L}$ be the set of lattices
\begin{equation}\label{eqn : lattice set}
\mathcal{L} = \left\{\Lambda_C : C \in \mathcal{B}\right\}.
\end{equation}
We now restate Minkowski-Hlawka Theorem proved in \cite{Loeliger} in a slightly different manner.

\begin{lem}[Minkowski-Hlawka Theorem]
Let $f$ be a Riemann integrable function $\mathbb{R}^n \rightarrow \mathbb{R}$ of bounded support. For any integer $k$,  $0 < k < n$ and any fixed $V$, let $\mathcal{B}$ be any balanced set of linear $(n,k)$ codes over $\mathbb{Z}_p$. As $p \rightarrow \infty$, $\gamma \rightarrow 0$ such that $\gamma^n p^{n-k} = V$, at least three-fourths of the lattices in the set $\mathcal{L}$ satisfy the following relationship
\begin{equation}\label{eqn : MH condition}
\sum_{v \in \gamma \Lambda_C : v \neq 0} f(v) \leq \frac{4}{V} \int_{\mathbb{R}^n} f(v)dv.
\end{equation}
\end{lem}
The proof of the lemma is exactly similar to the proof of \cite[Theorem 1]{Loeliger} with few minor modifications, and is omitted here.

We consider a single user point to point additive noise channel
\begin{equation}\label{eqn : Gaussian channel}
Y = X + Z
\end{equation}
where $X$ is the transmitted signal, $Y$ the received signal and $Z$ is the additive noise of zero mean and variance equal to $\sigma^2$ that corrupts the transmitted signal at the receiver. If the transmitted word over time is a lattice point, then it can be shown that a suitable lattice and a decoding strategy exists such that the probability of decoding error can be made arbitrarily small as the number of dimensions of the lattice increases. This result is stated formally in the following lemma.

\begin{lem}[\cite{Loeliger}]\label{lem : Loeliger}
Consider a single user point to point additive noise channel described in (\ref{eqn : Gaussian channel}). Let $\mathcal{B}$ be a balanced set of linear $(n,k)$ codes over $\mathbb{Z}_p$. Averaged over all lattices from the set $\mathcal{L}$ given in (\ref{eqn : lattice set}), each with a fundamental volume $V$, we have that for any $\delta > 0$, the average probability of decoding error is bounded by
\begin{equation}
\overline{P_e} < (1 + \delta) \frac{2^{n \frac{1}{2}\log(2\pi e \sigma^2)}}{V}.
\end{equation}
for sufficiently large $p$ and small $\gamma$ such that $\gamma^n p^{n-k} = V$. Hence, the probability of decoding error for at least three fourths of the lattices in $\mathcal{L}$ satisfies
\begin{equation}\label{eqn : Loeliger condition}
Pe < 4(1 + \delta) \frac{2^{n \frac{1}{2}\log(2\pi e \sigma^2)}}{V}.
\end{equation}
\end{lem}
The proof of the lemma is described in \cite{Loeliger} and is omitted here. The next Lemma summarizes the main result of \cite{UrbankeRimoldi}.

\begin{lem}\label{lem : Urbanke Rimoldi}
Consider a single user point to point additive noise channel in (\ref{eqn : Gaussian channel}) where the noise is AWGN with zero mean and variance equal to $\sigma^2$. Let $\Lambda$ be any lattice generated from Construction A that satisfies (\ref{eqn : MH condition}). Then, we can choose the fundamental volume of the lattice $V$, shift $s$ and a shaping region $S$ such that the lattice code $(\Lambda + s) \cap S$ achieves a rate $R$ with arbitrarily small average probability of error if
\begin{displaymath}
R \leq \frac{1}{2} \log\left(1 + \frac{P}{\sigma^2}\right).
\end{displaymath}
\end{lem}
The proof of the lemma is described in \cite{UrbankeRimoldi}.
In the next section, we show that for the three user symmetric Gaussian interference channel, all the users can achieve a symmetric rate of $\frac{1}{2}\log(1 + P)$ if the interference is ``very strong".

\section{Achievability Proof for Three User Symmetric Gaussian Interference Channel}\label{Sec:Achievability}
The transmitters employ lattice coding as a transmission strategy. In this section, we show that each user can achieve a symmetric rate $R < \frac{1}{2}\log(1 + P)$ under very strong interference condition.  As the channel is symmetric, we use the same lattice at each transmitter $\Lambda$ and is generated using construction A. We denote the Voronoi region of the lattice $\Lambda$ by $\Omega$ and the volume of the Voronoi region by $V$. The receivers first decode the total interference caused by other transmitters and then decode their message. Each transmitter uses a shift $s$ and a shaping region $\mathcal{S}$. Let $\mathcal{S}_1$ denote a $n$ dimensional sphere of radius $\sqrt{nP}$, and $\mathcal{S}_2$ denote a $n$ dimensional sphere of radius $\sqrt{nP'}$ where $P' < P$. Then the shaping region $\mathcal{S}$ is given by $\mathcal{S} = \mathcal{S}_1 \backslash\mathcal{S}_2$. Let $V_{\mathcal{S}}$ denote the volume of the shaping region $\mathcal{S}$. The codebook for each transmitter is given by $\mathcal{C} = (\Lambda + s) \cap \mathcal{S}$. The message set at each transmitter is denoted by $M = \{1, 2, \ldots, 2^{nR}\}$. For each message $m \in M$, the transmitter $i$ assigns a codeword $X_i(m) \in \mathcal{C}$.

We choose $R, R', P'$ and $P$ such that
\begin{displaymath}
R < R' < \frac{1}{2}\log(1 + P') < \frac{1}{2}\log(1 + P)
\end{displaymath}

We describe the decoding strategy for receiver 1 and the corresponding probability of error calculations. The analysis is similar for receivers $2$ and $3$ and is skipped here. We first describe the choice of lattice $\Lambda$ and the shift $s$. The lattice $\Lambda$ is chosen such that:
\begin{itemize}
\item Condition (\ref{eqn : MH condition}) (Minkowski-Hlawka condition) is satisfied.
\item The volume of the Voronoi region $V = 2^{-n R'} V_{\mathcal{S}}$.
\item In decoding the interference, the probability of error is upper bounded by (\ref{eqn : Loeliger condition}) with $\sigma^2 = 1 + P$.
\end{itemize}
We choose a shift $s$ such that the codebook $\left|\mathcal{C}\right| \geq 2^{nR}$. The existence of such a shift is guaranteed by \cite{UrbankeRimoldi} for large $n$.

Decoding Strategy for Receiver $1$: Receiver $1$ first cancels the sum of the interference caused by transmitters $2$ and $3$ and then decodes the message intended for it. The received output $Y_1$ is given by
\begin{displaymath}
Y_1 = X_1 + a X_2 + a X_3 + Z_1.
\end{displaymath}

As each transmitter uses the same lattice $\Lambda$, the interference caused by transmitters $2$ and $3$ at receiver $1$ is aligned and is an element of $a \Lambda$. Here, we use the fact that the receiver knows the shift $s$ used by transmitters $2$ and $3$ and cancels them out. We use the Loeliger framework in \cite{Loeliger} in decoding the total interference. The volume of the Voronoi region of the interference lattice is given by $a^n V$. The total noise seen in decoding the interference is given by
\begin{displaymath}
I_1 = X_1 + Z_1.
\end{displaymath}
The noise power is limited by $1 + P$. With the choice of our lattice, the probability of decoding error denoted by $P_{e,I}$ is upper bounded by
\begin{equation}\label{eqn : probability of interference error}
P_{e,I} < 4(1 + \delta) \frac{2^{n\frac{1}{2}\log(2\pi e (1+P))}}{a^n V}
\end{equation}
Hence, the probability of error decays if
\begin{equation}\label{eqn : condition for decay}
\frac{1}{2}\log\left(\frac{2\pi e (1 + P)}{a^2}\right) - \frac{1}{n}\log V < 0.
\end{equation}

Lemma \ref{lem : Loeliger} guarantees the choice of lattice $\Lambda$ such that (\ref{eqn : probability of interference error}) is satisfied. After decoding the total interference caused by transmitters $2$ and $3$, receiver $1$ decodes its message from the resulting point to point AWGN channel. In decoding its own message, receiver $1$ uses the nearest neighbor decoding approach as described in \cite{UrbankeRimoldi}. As the lattice $\Lambda$ satisfies (\ref{eqn : MH condition}), we can use the Urbanke - Rimoldi approach to decode the intended message at the receiver.

Then, from \cite{UrbankeRimoldi}, it follows that the average probability of decoding error decays with $n$.
Hence, receiver $1$ can decode its message successfully if
\begin{equation}\label{eqn : condition1}
R' < \frac{1}{2}\log(1 + P)
\end{equation}
Also by choosing sufficiently large $n$, the condition for decoding the interference with decaying probability of error as given in (\ref{eqn : condition for decay}) reduces to
\begin{equation}\label{eqn : condition2}
R' < \frac{1}{2}\log\left(\frac{a^2 P}{1 + P}\right).
\end{equation}

The very strong interference condition comes when the rate constraints imposed by decoding the interference is less binding than the constraint imposed by decoding their respective messages at the receivers. Hence, the very strong interference condition is given when the constraint on $R'$ due to (\ref{eqn : condition2}) is less binding than that due to (\ref{eqn : condition1}), or when
\begin{equation}
a^2 \geq \frac{(P+1)^2}{P}.
\end{equation}

By choosing $R'$ and $P'$ appropriately, we can show that user $1$ can achieve a rate arbitrarily close to $\frac{1}{2}\log(1 + P)$ under very strong interference condition. The decoding strategy for receivers $2$ and $3$ are similar, and lead to identical constraints on rates. Hence, users $2$ and $3$ can also achieve a rate arbitrarily close to $\frac{1}{2}\log(1 + P)$ when the interference is very strong. This completes the proof of Theorem $1$. The proof for any $K$ is similar to the one presented here for $K = 3$.

Three observations are made at this point.

Remark 1: In the proof of Theorem 1, we use a lattice generated using Construction A. We need to choose a lattice such that both (\ref{eqn : MH condition}) and (\ref{eqn : probability of interference error}) are satisfied. The existence of a lattice that satisfies both conditions can be seen from Lemmas 1 and 2. By choosing any balanced set of linear $(n,k)$ codes over $\mathbb{Z}_p$, three-fourth of the resulting set of lattices satisfies (\ref{eqn : MH condition}). Also, in Lemma 2, we show that for three-fourths of the lattices, (\ref{eqn : probability of interference error}) is satisfied. Hence, there exists at least one lattice such that both the conditions are satisfied.

Remark 2: In decoding the interference at each receiver, the number of interference points does not affect the decoding probability of error. The only condition we need for the probability of decoding error to decay is that the volume of the Voronoi region of the lattice be greater than the volume of a ``typical" noise set.

Remark on Theorem 2 : The proof of Theorem 2 follows along lines very similar to that of Theorem 1. The main difference is that in the proof of Theorem 2, we use the Loeliger approach in \cite{Loeliger} to decode the interference as well as the message. By using the Loeliger approach in decoding the message, we note that each user can achieve a rate of only $\frac{1}{2}\log(P)$.

\section{Conclusions}
In this paper, we derived a ``very strong" interference regime for the symmetric $K$ user Gaussian interference channel. That is, if the channel gain satisfies
\begin{displaymath}
a^2 \geq \frac{(P+1)^2}{P},
\end{displaymath}
the capacity region of the interference channel is as if no interference is present. Each receiver can decode the total interference seen, and then decode its message. The rate achieved by each user in this scenario is equal to $\frac{1}{2}\log(1 + P)$, which is the maximum rate that can be achieved when no interference is present. The condition for ``very strong" interference presented here in (\ref{eqn:tight constraint}) is much tighter than the natural extension of the ``very strong" interference condition for the two user interference channel (see (\ref{eqn:strongdecode})) in \cite{CarleialVeryStrongInterference}.
\vspace{0.35cm}
\section*{Acknowledgement}
We thank Bobak Nazer for useful discussions and comments.
\vspace{0.35cm}

\end{document}